\documentclass[copyright]{eptcs}
\providecommand{\event}{J.~Dassow, G.~Pighizzini, B.~Truthe (Eds.): 11th International Workshop\\
on Descriptional Complexity of Formal Systems (DCFS 2009)} 
\providecommand{\volume}{3}
\providecommand{\anno}{2009}
\providecommand{\firstpage}{137} 
\providecommand{\eid}{13}
\usepackage{breakurl}        

\input{dcfs.tex}

\usepackage{latexsym}
\usepackage{amsmath}\usepackage{amssymb}

\newcommand{\dfa}{\textsc{2dfa}}
\newcommand{\nfa}{\textsc{2nfa}}
\newcommand{\dsp}{\textsc{dspace}}
\newcommand{\nsp}{\textsc{nspace}}
\newcommand{\qeq}{\stackrel{{\scriptscriptstyle\rm ?}}{=}}
\newcommand{\qi}{q_{\scriptscriptstyle\rm I}}
\newcommand{\qf}{q_{\scriptscriptstyle\rm F}}
\newcommand{\qpi}{q'_{\scriptscriptstyle\rm I}}
\newcommand{\bull}{^{\bullet}}
\newcommand{\tast}{^{\ast}}
\newcommand{\tbox}{^{\scriptscriptstyle\Box}}

\newcommand{\set}[1]{\{{#1}\}}
\newcommand{\slen}[1]{|{#1}|}
\newcommand{\quo}[1]{``{#1}''}
\newcommand{\iit}[1]{\emph{\/(#1)\/}}

\newcommand{\bsection}[2]{\section[#1]{#1\label{#2}}\ignorespaces}

\newcommand{\btheorem}[1]{\begin{theorem}\label{#1}\ignorespaces}
\newcommand{\etheorem}{\end{theorem}}

\newcommand{\bcorollary}[1]{\begin{corollary}\label{#1}\ignorespaces}
\newcommand{\ecorollary}{\end{corollary}}

\newcommand{\bequations}{\begin{equation}\begin{array}{rcl}}
\newcommand{\eequations}[1]{\end{array}\label{#1}\end{equation}%
  \ignorespaces}

\newcommand*{\eitem}[1]{\item\label{#1}\ignorespaces}
\newcommand*{\benumerate}{\begin{enumerate}}
\newcommand*{\cenumerate}{\begin{enumerate}%
  \setcounter{enumi}{\value{lastenumi}}}
\newcommand*{\eenumerate}{\unskip
  \setcounter{lastenumi}{\value{enumi}}\end{enumerate}}
\newcounter{lastenumi}

\newcommand*{\bitemize}{\begin{itemize}}
\newcommand*{\eitemize}{\end{itemize}}

\newcommand*{\bproof}{\begin{proof}}
\newcommand*{\eproof}{\end{proof}}

\newcommand*{\bstatement}[1]{\paragraph*{#1.}\begingroup\itshape\ignorespaces}
\newcommand*{\estatement}{\unskip\endgroup\bigbreak}

\begin{document}%
\title{Translation from Classical Two-Way Automata
  to Pebble Two-Way Automata\,%
\thanks{Supported by the Slovak Grant Agen\-cy for Science (VEGA)
  under contract 1/0035/09.}}%
\def\titlerunning{Translation from Classical Two-Way Automata to Pebble Two-Way Automata}%
\author{\makebox[0pt][c]{Viliam Geffert}\hspace{5cm}%
\makebox[0pt][c]{L\kern-0.8ex\raise0.1ex\hbox{'}\kern0.1ex{}ubom\'{\i}ra
  I\v{s}to\v{n}ov\'{a}}
\institute{Department of Computer Science --
  P.\,J.~\v{S}af\'{a}rik University\\
  Jesenn\'{a} 5 -- 040\,01 Ko\v{s}ice -- Slovakia}
\email{\makebox[0pt][c]{viliam.geffert@upjs.sk}\hspace{5cm}%
\makebox[0pt][c]{lubomira.istonova@upjs.sk}}
}
\def\authorrunning{V.~Geffert, L.~I\v{s}to\v{n}ov\'{a}}
\maketitle
\begin{abstract}
We study the relation between the standard two-way automata and
more powerful devices, namely, two-way finite automata with an
additional \quo{pebble} movable along the input tape. Similarly as
in the case of the classical two-way machines, it is not known
whether there exists a polynomial trade-off, in the number of
states, between the nondeterministic and deterministic pebble
two-way automata. However, we show that these two machine models
are not independent: if there exists a polynomial trade-off for
the classical two-way automata, then there must also exist
a polynomial trade-off for the pebble two-way automata. Thus, we
have an upward collapse (or a downward separation) {}from the
classical two-way automata to more powerful pebble automata, still
staying within the class of regular languages. The same upward
collapse holds for complementation of nondeterministic two-way
machines.

These results are obtained by showing that each pebble machine can
be, by using suitable inputs, simulated by a classical two-way
automaton with a linear number of states (and vice versa), despite
the existing exponential blow-up between the classical and pebble
two-way machines.
\end{abstract}

\bsection{Introduction}{s:int}
Relation between determinism and nondeterminism is one of the key
topics in theoretical computer science. The most famous is the
$\textsc{P}\qeq\textsc{NP}$ question, but the oldest problem of
this kind is\linebreak $\dsp(n)\qeq\nsp(n)$. Similarly, we do not know
whether $\dsp(\log n)\qeq\nsp(\log n)$. However, a positive
answer for the $O(\log n)$ space would imply the positive answer
for the $O(n)$ space, and hence the answers to these two questions
are not independent. Analogically, a collapse for the
$O(\log\log n)$ space would imply the same collapse for the
$O(\log n)$ space. (For a survey and bibliography about such
translations, see e.\,g.~\cite{Ge98,Sz94b}.) Analogous upward
translations can be derived for time complexity classes.

\smallbreak
At first glance, the problem has been resolved for finite state
automata. Even a two-way nondeterministic finite automaton (\nfa,
for short) and hence any simpler device as well (e.\,g., its
deterministic version, \dfa) can recognize a regular language
only. Thus, \nfa's can be converted into deterministic one-way
automata. However, the problem reappears, if we take into account
the size of these automata, measured in the number of states.

On one hand, eliminating nondeterminism in one-way $n$-state
automata does not cost more than $2^n$ states (by the classical
subset construction), and there exist witness regular languages
for which exactly $2^n$~states are indeed required.

On the other hand, we know very little about eliminating
nondeterminism in the two-way case: it was conjectured by Sakoda
and Sipser~\cite{SS78} that there must exist an exponential
blow-up for the conversion of \nfa's into \dfa's. Nevertheless,
the best known lower bound is~$\Omega(n^2)$~\cite{Ch86}, while the
best conversion uses about $2^{n^{2}}$ states (converting
actually into deterministic one-way machines). Thus, it is not
clear whether there exists a polynomial trade-off. The problem has
been attacked several times by proving exponential lower bounds
for restricted versions of \dfa's: Sipser~\cite{Si79}\,---\,for
sweeping machines (changing the direction of the input head
movement at the endmarkers only); Hromkovi\v{c} and
Schnitger~\cite{HS03}\,---\,for oblivious machines (moving the
input head along the same trajectory on all inputs of the same
length); Kapoutsis~\cite{Ka07}\,---\,a computability separation
for \quo{moles} (seeing only a part of the input symbol thus
traveling \quo{in a network of tunnels} along the input). For
machines accepting unary languages, a subexponential upper bound
$2^{O(\log^{2}n)}$ has been obtained~\cite{GMP03}.

It was even observed~\cite{SS78} that there exists a family of
regular languages $\set{B_n:n\ge 1}$ which is \emph{complete} for
the two-way automata, playing the same role as, e.\,g., the
satisfiability of boolean formulas for the
$\textsc{P}\qeq\textsc{NP}$ question or the reachability in graphs
for $\dsp(\log n)\qeq\nsp(\log n)$: the trade-off between the
\nfa's and \dfa's is polynomial if and only if it is polynomial
for~$B_n$, i.\,e., if and only if~$B_n$~can be accepted by a \dfa\
with a polynomial number of states. (For \nfa's, \,$n$~states are
enough to accept~$B_n$.)

\smallbreak
In the absence of a solution for the general case, it is quite
natural to ask whether some properties of the two-way automata
cannot be translated into more powerful machines, in perfect
analogy with the corresponding results for the upward translation
established for the classical space and time complexity classes.
So far, the only result of this kind~\cite{BL77} is that if an
exponential trade-off between \nfa's and \dfa's could be obtained
already by using a subset of the original language that consists
of polynomially long strings, then
$\dsp(\log n)\ne\nsp(\log n)$.

\smallbreak
In the same spirit, we shall study the relation between the
standard two-way automata and more powerful devices, namely,
two-way nondeterministic and deterministic finite automata
equipped with a single additional \quo{pebble}, movable along
the input tape (pebble-\nfa, pebble-\dfa, respectively). Despite
the fact that such pebble can be used to mark some input tape
position, even a nondeterministic pebble machine cannot accept
a nonregular language~\cite{CIPR86,Sz94b}. However, measured in
the number of states, the pebble machines are much more powerful.
Converting a pebble-\dfa\ to a classical \nfa\ may require an
expo\-nen\-tial blow-up, i.\,e., the loss of the pebble cannot be
compensated economically by gaining nondeterminism. (See
Theorem~\ref{t:blow} below.)

Similarly as in the case of the classical machines, we do not know
whether there exists a polynomial trade-off between the
pebble-\nfa's and pebble-\dfa's. However, we shall show that
these two models are related: if there exists a polynomial
transformation {}from the classical \nfa's to \dfa's, then there
must also exist a polynomial transformation, with the same degree
of the polynomial, {}from the pebble-\nfa's to pebble-\dfa's.
Thus, we have an upward collapse (and a downward separation)
between the classical two-way automata and the much more powerful
pebble model, within the class of regular languages.

A similar upward collapse holds for the trade-off between
a two-way nondeterministic automaton accepting a language~$L$ and
a machine for the complement of~$L$: if the trade-off is
polynomial for the classical \nfa's, it must also be polynomial
for the pebble-\nfa's. (The problem is open for both these
models.)

These results are obtained by showing that each pebble-\nfa\ (or
pebble-\dfa) can be, by using suitable inputs, simulated by
a classical \nfa\ (or \dfa, respectively) with only a linear
number of states, despite the existing exponential blow-up between
the classical and pebble machines. The same holds for the
corresponding conversions {}from the classical machines to pebble
machines.

\bsection{Preliminaries}{s:pre}
Here we introduce some basic notation and properties for the
computational models we shall be dealing with. For a more detailed
exposition and bibliography related to regular languages, the
reader is referred to~\cite{HMU01,Ka07,MP01}.

\smallbreak
A \emph{two-way nondeterministic finite automaton} (\nfa, for
short) is a quintuple $A=(Q,\Sigma,\delta,\qi,F)$, in which
$Q$~is the finite set of states, $\Sigma$~is the finite input
alphabet,
$\delta:Q\times(\Sigma\cup\set{\vdash,\dashv})\rightarrow
2^{Q\times\set{-1,0,+1}}$
is the transition function, $\vdash,\dashv\;\not\in\Sigma$ are two
special symbols, called the left and the right endmarker,
respectively, $\qi\in Q$ is the initial state, and $F\subseteq Q$
is the set of accepting (final) states.

The input is stored on the input tape surrounded by the two
endmarkers. In one move, $A$~reads an input symbol, changes its
state, and moves the input head one cell to the right, left, or
keeps it stationary, depending on whether $\delta$ returns $+1$,
$-1$, or~$0$, respectively. The input head cannot move outside the
zone delimited by the endmarkers: transitions in the form
$\delta(q,\dashv)\ni(q',+1)$ or $\delta(q,\vdash)\ni(q',-1)$
are not allowed. If $\slen{\delta(q,a)}=0$, the machine
\emph{halts}.

The machine accepts the input, if there exists a computation path
starting in the initial state~$\qi$ with the head on the left
endmarker and reaching, anywhere along the input tape, an
accepting state $q\in F$.

The automaton $A$ is said to be \emph{deterministic} (\dfa),
whenever $\slen{\delta(q,a)}\le 1$, for any $q\in Q$ and
$a\in\Sigma\cup\set{\vdash,\dashv}$.

\smallbreak
We also study a more powerful model, namely, a two-way finite
automaton equipped with an additional \quo{pebble} placed on the
input tape. The action of the pebble machine depends on the
current state, the currently scanned input tape symbol, and the
presence of the pebble on this symbol. The action consists of
changing the current state, moving the input head and, optionally,
if the pebble is placed on the current symbol, moving also the
pebble in the same direction.

Formally, a \emph{one-pebble two-way nondeterministic finite
automaton} (pebble-\nfa, for short) is\linebreak
\hbox{$A=(Q,\Sigma,\delta,\qi,F)$}, where $Q,\Sigma,\qi,F$ are defined as
above, but the transition function is of the form
$\delta:Q\times(\Sigma\cup\Sigma\bull\cup
\set{\vdash,\dashv,\vdash\bull,\dashv\bull})\rightarrow
2^{Q\times\set{-1,0,+1,-1\bull,+1\bull}}$.
The presence of the pebble on the current input tape symbol
$a\in\Sigma\cup\set{\vdash,\dashv}$ is indicated by using
$a\bull\in\Sigma\bull\cup\set{\vdash\bull,\dashv\bull}$, while the
new input head movements $-1\bull,+1\bull$ are introduced to move
the pebble. More precisely, a classical transition
$\delta(q,a)\ni(q',d)$, with
$a\in\Sigma\cup\set{\vdash,\dashv}$ and $d\in\set{-1,0,+1}$, is
applicable only if the pebble in not placed on the current input
tape symbol (change the current state {}from~$q$ to~$q'$ and move
the input head in the direction~$d$), while
$\delta(q,a\bull)\ni(q',d)$ can be executed only if the pebble
is placed on $a\in\Sigma\cup\set{\vdash,\dashv}$ at the moment
(move the input head in the direction~$d$, but leave the pebble in
its original position). Finally,
$\delta(q,a\bull)\ni(q',d\bull)$, with
$d\bull\in\set{-1\bull,+1\bull}$, moves also the pebble in the
same direction~$d$, together with the input head. Transitions in
the form $\delta(q,a)\ni(q',d\bull)$ are meaningless, and hence
not allowed.

The machine $A$ starts its computation in the initial state~$\qi$
with both the input head and the pebble placed on the left
endmarker, and accepts by reaching, anywhere along the input tape,
a final state $q\in F$. Similarly, the final position of the
pebble is irrelevant for acceptance. A one-pebble two-way
\emph{deterministic} finite automaton (pebble-\dfa) is defined in
the usual way.

\medbreak
It is known~\cite{CIPR86} (see also Theorem~15.3.5
in~\cite{Sz94b}) that even nondeterministic Turing machines
equip\-ped with a single pebble and a worktape space of size
$o(\log\log n)$ can accept regular languages only. Since
pebble-\nfa's may be viewed as one-pebble Turing machines with
$O(1)$ worktape space, all models introduced above (\dfa, \nfa,
pebble-\dfa, pebble-\nfa) share the same expressive
power\,---\,they all recognize the same class of regular
languages.

However, if we take into account their number of states, the power
is different. Converting a pebble-\dfa\ to a classical \nfa\ may
require an expo\-nen\-tial blow-up. That is, the loss of the
pebble cannot be paid by gaining nondeterminism.

\btheorem{t:blow}
For each $m\ge 1$, there exists a finite unary language~$L_m$ that
can be accepted by a pebble-\dfa\ with $O(m^2\cdot\log m)$
states, but for which each \nfa\ requires at least
$2^{\Omega(m\cdot\log m)}$ states.
\etheorem
\bproof
Let $M=p_1\cdotp_2\cdot\ldots\cdotp_m$, where
$p_i$~denotes the $i$-th prime number, and let
$L_m=\set{1^{\ell}:\ell<M}$.

The pebble machine~$A$ recognizing~$L_m$ utilizes the fact that
$\ell<M$ if and only if no $x\in\set{1,\ldots,\ell}$ is a common
multiple of $p_1,p_2,\ldots,p_m$. Therefore, $A$~repeatedly
checks, for $x=1,\ldots,\ell$, if $x$~is divisible by the primes
$p_1,p_2,\ldots,p_m$. The value of~$x$ is represented by the
distance of the pebble {}from the left endmarker. In order to
check if $p_i$ divides~$x$, $A$~traverses between the pebble
position and the left endmarker and counts modulo~$p_i$
(alternating right-to-left traversals with left-to-right
traversals for odd/even values of~$i$). If $A$~finds
a prime~$p_i$ not dividing~$x$, it does not check the next
prime~$p_{i+1}$ but, rather, enters the initial state~$\qi$ in
which it searches for the pebble and then moves the pebble one
position to the right. After that, $A$~can start checking the next
value of~$x$ for divisibility by $p_1,p_2,\ldots,p_m$ or, if the
pebble has reached the right endmarker, $A$~can halt in an
accepting state~$\qf$. Carefully implemented, $A$~uses only
$2+p_1+p_2+\dots+p_m$ states. By the Prime Number
Theorem (see, e.\,g.,~\cite{EE85}), we have
$p_i=(1+o(1))\cdot i\cdot\ln i$, which gives
$2+p_1+p_2+\dots+p_m\le O(m^2\cdot\log m)$.

On the other hand, each classical \nfa~$A'$ recognizing~$L_m$ must
use at least $M-1$ states. This can be seen by the use of
$n\rightarrow n+n!$ method~\cite{CHR91,Ge91}: On the
input~$1^{M-1}$, a machine~$A'$ with fewer states than
$M-1$ cannot traverse the input tape {}from left to right
without going into a loop, i.\,e., without repeating the same state
after traveling some $h$~positions to the right, where
$h\le M-1$. Thus, by iterating this loop
$(M-1)!/h=\prod_{i=1,i\ne h}^{M-1}{i}$ \,more times, we get
a valid computation path traversing $(M-1)+(M-1)!$
positions to the right. Therefore, if $A'$~can get {}from a state
$q_1$ to~$q_2$ by traversing the entire input~$1^{M-1}$, it can
also get {}from $q_1$ to~$q_2$ by traversing the entire input
$1^{(M-1)+(M-1)!}$. The same holds for right-to-left
traversals and also for U-turns, i.\,e., for computations starting
and ending at the same endmarker. Thus, by induction on the number
of visits at the endmarkers, we get that if $A'$~accepts the
input~$1^{M-1}$, it must also accept the input
$1^{(M-1)+(M-1)!}$, which is a contradiction. Therefore, each
\nfa~$A'$ recognizing~$L_m$ must use at least
$M-1=p_1\cdotp_2\cdot\ldots\cdotp_m-1$ states.
Since
$p_1\cdotp_2\cdot\ldots\cdotp_m\ge m^{\Omega(m)}$
(see, e.\,g., Lemma~4.14 in~\cite{Ge98}), we have
$M-1\ge 2^{\Omega(m\cdot\log m)}$.
\eproof

\bsection{Translation}{s:tra}
In this section, we first show that each pebble-\nfa~$M$ (or
pebble-\dfa) can be, in a way, using a suitable encoding of the
original input, simulated by a \nfa~$M'$ (or \dfa, respectively)
without a pebble. Then we show the corresponding conversions in
the opposite direction, {}from the classical two-way machines to
two-way machines with a pebble. The cost, in the number of states,
will be linear for all these conversions, despite the exponential
blow-up presented by Theorem~\ref{t:blow}. After that, we shall
derive some consequences of these translations.

\smallbreak
In what follows, we shall need a function~$P$ that maps each
input~$w$ of the given pebble automaton~$M$ into a new
word~$P(w)$ providing all possible positions of the pebble
in~$w$. This image~$P(w)$ can be used as an input for
a classical automaton~$M'$ (no pebble), such that $M'$
accepts~$P(w)$ if and only if $M$ accepts~$w$. Let
$P:\Sigma\tast\rightarrow
(\Sigma\cup
\set{\triangleright,\triangleleft}\cup\Sigma\tbox)\tast$
maps a word $w=a_1\ldots a_k$ as follows:
\bequations
  P(a_1\ldots a_k) &=&
    a_1\ldots a_k\triangleleft
    \triangleright a_1\tbox a_2\ldots a_k\triangleleft
    \triangleright a_1 a_2\tbox\ldots a_k\triangleleft
    \triangleright\ldots\\
  &&\ldots\triangleleft
    \triangleright a_1\ldots a_{k-1}\tbox a_k\triangleleft
    \triangleright a_1\ldots a_{k-1} a_k\tbox\triangleleft
    \triangleright a_1\ldots a_k\,,
\eequations{e:map}
where $\triangleright,\triangleleft$ are new symbols and
$\Sigma\tbox=\set{a\tbox:a\in\Sigma}$, that is,
$\Sigma\tbox$~simply denotes the letters of the original alphabet
marked by some box.

Thus, $P(a_1\ldots a_k)$ consists of $k+2$ segments,
enumerated {}from~$0$. The $p$-th segment, for\linebreak
$p=0,\ldots,k+1$, will be used by~$M'$ to simulate~$M$ in
situations when $M$~has the pebble placed on the $p$-th position
of the input tape. For these reasons, the $p$-th segment is of
type $\triangleright a_1\ldots a_p\tbox\ldots a_k\triangleleft$,
that is, the $p$-th symbol is marked by the box. (Except for
$p=0$ and $p=k+1$, there is exactly one such \quo{pseudo
pebble} in each segment.) The symbols $\triangleright$
and~$\triangleleft$ are \quo{stoppers}, imitating the left and
right endmarkers of the original input tape. The first and last
segments are of special kind, representing the situations when
$M$~has the pebble placed on the left or right endmarker,
respectively, with no letters marked by the box.

As an example, if $w=a_1a_2a_3$, then
$P(a_1a_2a_3)=
a_1a_2a_3\triangleleft
\triangleright a_1\tbox a_2a_3\triangleleft
\triangleright a_1a_2\tbox a_3\triangleleft
\triangleright a_1a_2a_3\tbox\triangleleft
\triangleright a_1a_2a_3$.
Thus, taking also into account the endmarkers, the input tape for
the pebble automaton~$M$ is in the form
$\vdash a_1a_2a_3\dashv$ while the input tape for a classical
automaton~$M'$ (no pebble) in the form
$$\vdash a_1a_2a_3\triangleleft
\triangleright a_1\tbox a_2a_3\triangleleft
\triangleright a_1a_2\tbox a_3\triangleleft
\triangleright a_1a_2a_3\tbox\triangleleft
\triangleright a_1a_2a_3\dashv.$$
Similarly, for $w=\varepsilon$, we have
$P(\varepsilon)=
\varepsilon\triangleleft\triangleright\varepsilon=
\triangleleft\,\triangleright$,
that is, the input tapes for $M$ and~$M'$ are $\vdash\,\dashv$ and
$\vdash\triangleleft\,\triangleright\dashv$, respectively.
Therefore, the left and right endmarkers can be handled by~$M'$ as
if marked by the \quo{pseudo pebble} box, that is, the symbols
$\vdash,\dashv$ can be viewed as if equal to
$\vdash\tbox,\dashv\tbox$, respectively.

\btheorem{t:up}
\iit{a}~For each pebble-\nfa\ $M=(Q,\Sigma,\delta,\qi,F)$ with
$m$~states, there exists a classical \nfa\
$M'=(Q',\Sigma',\delta',\qpi,F')$ with at most
$3\cdot m$ states such that, for each input
$w\in\Sigma\tast$, $M'$~accepts\linebreak
\hbox{$w'=P(w)\in{\Sigma'}\tast$} if
and only if $M$ accepts~$w$. Here
$\Sigma'=\Sigma\cup
\set{\triangleright,\triangleleft}\cup\Sigma\tbox$
and $P$~is the mapping function defined by~{\rm(\ref{e:map})}.

\iit{b}~Moreover, if $M$~is deterministic, then so is~$M'$.
\etheorem
\bproof
Note that $M'$ does not have to check whether its input
$w'\in{\Sigma'}\tast$ is indeed a valid image obtained by the use
of~$P$, i.\,e., if $w'=P(w)$, for some $w\in\Sigma\tast$.
Assuming that $w'=P(w)$, $M'$~simply checks whether $M$
accepts~$w$. If this assumption is wrong, the answer of~$M'$ can
be quite arbitrary. In general, an input $w'\in{\Sigma'}\tast$
does not necessarily have the structure described
by~(\ref{e:map}), for any $w\in\Sigma\tast$.

The basic idea is as follows. If, during the simulation, $M$~has
its pebble placed on the $p$-th position of~$w$, $M'$~works within
the $p$-th segment of~$P(w)$. The simulation is quite
straightforward and $M'$ does not have to leave this segment until
the moment when $M$~moves its pebble. Recall that $M'$~relies on
the assumption that the $p$-th segment contains one exact copy
of~$w$, correctly enclosed in between the symbols
$\triangleright$ and~$\triangleleft$, and that the current pebble
position of~$M$ is clearly marked inside this segment, i.\,e., there
is exactly one symbol marked with the box, namely, the symbol on
the $p$-th position. If this never-verified assumption were wrong,
the simulation could turn out to be wrong. Using this idea, we
start our construction of~$\delta'$, the transition function for
the automaton~$M'$, as follows.

\benumerate
  \eitem{i:up:dm} If $\delta(q,a)\ni(q',d)$, for some
    $q,q'\in Q$, $a\in\Sigma$, and $d\in\set{-1,0,+1}$, then
    $\delta'(q,a)\ni(q',d)$.
  \eitem{i:up:dl} If $\delta(q,\vdash)\ni(q',d)$, for some
    $q,q'\in Q$ and $d\in\set{0,+1}$, then
    $\delta'(q,\triangleright)\ni(q',d)$.
  \eitem{i:up:dr} If $\delta(q,\dashv)\ni(q',d)$, for some
    $q,q'\in Q$ and $d\in\set{-1,0}$, then
    $\delta'(q,\triangleleft)\ni(q',d)$.
  \eitem{i:up:pm} If $\delta(q,a\bull)\ni(q',d)$, for some
    $q,q'\in Q$, $a\in\Sigma$, and $d\in\set{-1,0,+1}$, then
    $\delta'(q,a\tbox)\ni(q',d)$.
  \eitem{i:up:pl} If $\delta(q,\vdash\bull)\ni(q',d)$, for some
    $q,q'\in Q$ and $d\in\set{0,+1}$, then
    $\delta'(q,\vdash)\ni(q',d)$.
  \eitem{i:up:pr} If $\delta(q,\dashv\bull)\ni(q',d)$, for some
    $q,q'\in Q$ and $d\in\set{-1,0}$, then
    $\delta'(q,\dashv)\ni(q',d)$.
\eenumerate

As soon as $M$~moves its pebble {}from the $p$-th position to the
right, $M'$~has to travel {}from the $p$-th segment to the next,
i.\,e., the $(p+1)$-st segment, and find the symbol marked by
the box within this segment. Assuming that the input is in the
form $w'=P(w)$, for some $w\in\Sigma\tast$, this only requires
to find the next symbol marked by the box lying to the right of
the current input position. Recall that the $(p+1)$-st segment
has, by assumption, the same structure; the only difference is in
the position of the symbol marked with the box, which corresponds
exactly to the changed position of the pebble for~$M$. Thus,
after finding the marked symbol within the neighboring segment,
$M'$~can resume the simulation.

\cenumerate
  \eitem{i:up:ppmr} If $\delta(q,a\bull)\ni(q',+1\bull)$, for
    some $q,q'\in Q$ and $a\in\Sigma$, we add the following
    instructions:
    \bitemize
      \item $\delta'(q,a\tbox)\ni(q'_{+1},+1)$, where
        $q'_{+1}$~is a passing-through state\,---\,a new copy
        of~$q'$,
      \item $\delta'(q'_{+1},x)\ni(q'_{+1},+1)$, for each
        $x\in\Sigma\cup\set{\triangleright,\triangleleft}$,
      \item $\delta'(q'_{+1},x\tbox) \ni(q',0)$, for each
        $x\tbox\in\Sigma\tbox\cup\set{\dashv}$.
    \eitemize
  \eitem{i:up:pplr} If $\delta(q,\vdash\bull)\ni(q',+1\bull)$,
    for some $q,q'\in Q$, then
    \bitemize
      \item $\delta'(q,\vdash)\ni(q'_{+1},+1)$.
      \item Transitions for~$q'_{+1}$ are defined in the same way
        as in the item~(\ref{i:up:ppmr}).
    \eitemize
\eenumerate

If $M$ moves the pebble to the left, $M'$~travels to the previous,
i.\,e., the $(p-1)$-st segment. This is resolved symmetrically
with the items~(\ref{i:up:ppmr}) and~(\ref{i:up:pplr}), replacing,
respectively, $\dashv,\vdash$ and~$+1$ by $\vdash,\dashv$
and~$-1$, thus using~$q'_{-1}$ (another passing-through copy of
the state~$q'$) instead of~$q'_{+1}$.

{}From the above construction, we get
$Q'=Q\cup Q_{+1}\cup Q_{-1}$, where $Q$~is the set of the original
states in~$M$ and $Q_{+1},Q_{-1}$ represent the sets of two new
copies of states in~$Q$, used for traversing to the neighboring
segments, introduced as $q'_{+1}$ and~$q'_{-1}$. The initial
state and the final states do not change: $\qpi=\qi$ and
$F'=F$. This completes the definition of~$M'$. It is also
easy to see that the construction preserves determinism.

\bstatement{Claim}
On the input~$w$, $M$~can get {}from its initial configuration,
i.\,e., {}from the state~$\qi$ with both the input head and the
pebble at the left endmarker, to a state $q\in Q$ with the input
head at a position~$h$ and the pebble at a position~$p$ \emph{if
and only if}, on the input~$P(w)$, $M'$~can get {}from its initial
configuration, i.\,e., {}from the state~$\qpi$ with the input head
at the left endmarker, to the same state $q\in Q$ with the input
head at the $h$-th position of the $p$-th segment.
\estatement

The argument for the \quo{$\Rightarrow$} part is shown by
induction on the number of computation steps executed by~$M$,
while the \quo{$\Leftarrow$} part uses an induction on the number
of times the machine $M'$~is in a state $q\in Q$, i.\,e., not in
a passing-through state $q\in Q_{+1}\cup Q_{-1}$ (instead of
induction on single computation steps).

As a consequence of this Claim, $M$~has an accepting computation
path on the input $w\in\Sigma\tast$, i.\,e., $M$~can reach an
accepting state~$q$ on the input~$w$ if and only if~$M'$, on the
input~$P(w)$, can reach the same state $q\in F'=F$ (not
a passing-through state in~$M'$), i.\,e., if and only if $M'$~has an
accepting computation path on the input
$w'=P(w)\in{\Sigma'}\tast$.
\eproof

Now we shall show a linear translation in the opposite direction.

\btheorem{t:dw}
\iit{a}~For each classical \nfa\ $N=(Q,\Sigma',\delta,\qi,F)$
with $n$~states, there exists a pebble-\nfa\
$N'=(Q',\Sigma,\delta',\qpi,F')$ with at most
\mbox{$5\cdot n$} states such that, for each input
$w\in\Sigma\tast$, $N'$~accepts~$w$ if and only if $N$~accepts
$w'=P(w)\in{\Sigma'}\tast$. Here
$\Sigma'=\Sigma\cup
\set{\triangleright,\triangleleft}\cup\Sigma\tbox$
and $P$~is the mapping function defined by~(\ref{e:map}).

\iit{b}~Moreover, if $N$~is deterministic, then so is~$N'$.
\etheorem
\bproof
Note that $N'$ does not have to be capable of simulating~$N$ on
all strings $w'\in{\Sigma'}\tast$. $N'$~simulates~$N$ only on
inputs in the form $w'=P(w)$, where $w\in\Sigma\tast$ is its own
input. Thus, $N'$~can utilize the fact that the string $P(w)$ has
the structure described by~(\ref{e:map}).

While $N$~works within the same segment of~$P(w)$, the simulation
by~$N'$ is quite straightforward: the endmarkers $\vdash,\dashv$
surrounding~$w$ are interpreted as stoppers
$\triangleright,\triangleleft$ in~$P(w)$, and the presence of the
pebble on a symbol $a\in\Sigma\cup\set{\vdash,\dashv}$ scanned by
the input head of~$N'$ indicates that $N$~reads
$a\tbox\in\Sigma\tbox$ or the corresponding endmarker
$\vdash,\dashv$. Thus, the pebble placed at a position~$p$
reflects the fact that $N'$~simulates, at the moment, $N$~working
within the $p$-th segment of~$P(w)$:

\benumerate
  \eitem{i:dw:sm} If $\delta(q,a)\ni(q',d)$, for some
    $q,q'\in Q$, $a\in\Sigma$, and $d\in\set{-1,0,+1}$, then
    $\delta'(q,a)\ni(q',d)$.
  \eitem{i:dw:sl} If $\delta(q,\triangleright)\ni(q',d)$, for
    some $q,q'\in Q$ and $d\in\set{0,+1}$, then
    $\delta'(q,\vdash)\ni(q',d)$.
  \eitem{i:dw:sr} If $\delta(q,\triangleleft)\ni(q',d)$, for
    some $q,q'\in Q$ and $d\in\set{-1,0}$, then
    $\delta'(q,\dashv)\ni(q',d)$.
  \eitem{i:dw:pm} If $\delta(q,a\tbox)\ni(q',d)$, for some
    $q,q'\in Q$, $a\in\Sigma$, and $d\in\set{-1,0,+1}$, then
    $\delta'(q,a\bull)\ni(q',d)$.
  \eitem{i:dw:pl} If $\delta(q,\vdash)\ni(q',d)$, for some
    $q,q'\in Q$ and $d\in\set{0,+1}$, then
    $\delta'(q,\vdash\bull)\ni(q',d)$.
  \eitem{i:dw:pr} If $\delta(q,\dashv)\ni(q',d)$, for some
    $q,q'\in Q$ and $d\in\set{-1,0}$, then
    $\delta'(q,\dashv\bull)\ni(q',d)$.
\eenumerate

Each time $N$~leaves the current segment, e.\,g., if it moves its
input head {}from the symbol~$\triangleleft$ to the right (that
is, in the next step, $N$~will read the symbol~$\triangleright$
belonging to the next segment), $N'$~does not try to move its
input head to the right {}from the right endmarker but, rather, it
temporarily interrupts the simulation and enters
a passing-through routine in which it traverses the entire
input~$w$ {}from right to left and, during this traversal, it
moves the pebble one position to the right. After that, with the
input head at the left endmarker of~$w$, $N'$~is ready to resume
the simulation on the next segment of~$P(w)$. Note that the
instructions defined in the item~(\ref{i:dw:cr}) cover also three
special subcases, namely, migration of the pebble {}from the left
endmarker to the first input symbol, {}from the last input symbol
to the right endmarker, or, for $w=\varepsilon$, {}from the left
endmarker directly to the right endmarker:

\cenumerate
  \eitem{i:dw:cr} If $\delta(q,\triangleleft)\ni(q',+1)$, for
    some $q,q'\in Q$, we add the following instructions:
    \bitemize
      \item $\delta'(q,\dashv)\ni(q'_{-1},-1)$, where $q'_{-1}$~is
        a new copy of~$q'$\,---\,a passing-through state
        searching for the pebble to the left,
      \item $\delta'(q'_{-1},x)\ni(q'_{-1},-1)$, for each
        $x\in\Sigma$,
      \item $\delta'(q'_{-1},x\bull)\ni(q'_{-2},+1\bull)$, for
        each $x\in\Sigma\cup\set{\vdash}$, where $q'_{-2}$~is
        another new copy of~$q'$\,---\,a passing-through state
        searching for the left endmarker,
      \item $\delta'(q'_{-2},x\bull)\ni(q'_{-2},-1)$, for each
        $x\in\Sigma\cup\set{\dashv}$,
      \item $\delta'(q'_{-2},x)\ni(q'_{-2},-1)$, for each
        $x\in\Sigma$,
      \item $\delta'(q'_{-2},\vdash)\ni(q',0)$.
    \eitemize
\eenumerate

Similarly, each time $N$~leaves the current segment for the
previous segment by moving {}from the symbol~$\triangleright$ to
the left (after which it will read~$\triangleleft$),
$N'$~interrupts the simulation and enters a routine traversing the
entire input~$w$ {}from left to right and, during this traversal,
it moves the pebble one position to the left. This is resolved
symmetrically with the item~(\ref{i:dw:cr}), replacing,
respectively, $\triangleleft,\dashv,\vdash$ and~$+1,-1$ by
$\triangleright,\vdash,\dashv$ and~$-1,+1$, thus
using~$q'_{+1},q'_{+2}$, some passing-through counterparts of
$q'_{-1},q'_{-2}$.

Thus, $Q'=Q\cup Q_{-1}\cup Q_{-2}\cup Q_{+1}\cup Q_{+2}$, where
$Q$~is the original set of states and
$Q_{-1}$, $Q_{-2}$, $Q_{+1}$, $Q_{+2}$ are four new copies of~$Q$.
Finally, $\qpi=\qi$ and $F'=F$.

\smallbreak
The argument showing that $N'$~accepts $w\in\Sigma\tast$ if and
only if $N$~accepts $P(w)\in{\Sigma'}\tast$ is very similar to
that of Theorem~\ref{t:up}: by induction on the number of steps
executed by~$N$ and by induction on the number of times $N'$~is in
a state $q\in Q$ (i.\,e., not in a passing-through state), we can
prove the claim saying that, on the input~$w$, $N'$~can reach
a state $q\in Q$ with the input head at a position~$h$ and the
pebble at a position~$p$ \emph{if and only if}, on the
input~$P(w)$, $N$~can reach the same state~$q$ with the input head
at the $h$-th position of the $p$-th segment. It is also easy to
see that the construction preserves determinism.
\eproof

Now we are ready to draw some consequences of the above
translations.

\btheorem{t:det}
If, for some function~$f(n)$, each \nfa\ with $n$~states can be
replaced by an equivalent \dfa\ with at most $f(n)$~states (no
pebbles), then each pebble-\nfa\ with $m$~states can be replaced
by an equivalent pebble-\dfa\ having no more than
$5\cdot f(3m)$ states.

In particular, if $f(n)\le O(n^k)$, that is, if there exists
a polynomial transformation {}from nondeterministic to
deterministic classical two-way automata, then there must also
exist a polynomial transformation, with the same degree of the
polynomial, {}from nondeterministic to deterministic two-way
automata equipped with a pebble, since
$5\cdot(3m)^k=(5\cdot3^k)\cdot m^k\le O(m^k)$.
\etheorem
\bproof
By Theorem~\ref{t:up}(a), each pebble-\nfa\ $M$ with $m$~states
accepting a language $L\subseteq\Sigma\tast$ can be replaced by
a classical \nfa~$M'$ with at most $3\cdot m$ states, accepting
some other language $L'\subseteq{\Sigma'}\tast$. However, for
each input $w\in\Sigma\tast$, $M$~accepts~$w$ if and only if
$M'$~accepts $P(w)\in{\Sigma'}\tast$. Here $P$~denotes the
mapping function defined by~(\ref{e:map}). By assumption,
$M'$~can be replaced by a classical \dfa~$N$, with at most
$f(3m)$~states, equivalent to~$M'$. Among others,
$M'$~accepts~$P(w)$ if and only if $N$~accepts~$P(w)$. Now, by
Theorem~\ref{t:dw}(b), we can replace~$N$ by a pebble-\dfa\ $N'$
with no more than $5\cdot f(3m)$ states, such that
$N$~accepts $P(w)\in{\Sigma'}\tast$ if and only if $N'$~accepts
$w\in\Sigma\tast$. Thus, for each input $w\in\Sigma\tast$,
$M$~accepts~$w$ if and only if $N'$~accepts~$w$, and hence these
two pebble machines are equivalent.%
\eproof

The situation for complementing nondeterministic machines is very
similar.

\btheorem{t:com}
If, for some function~$f(n)$, each \nfa\ with $n$~states can be
replaced by a \nfa\ with at most $f(n)$~states recognizing the
complement of the original language (no pebbles), then each
pebble-\nfa\ with $m$~states can be replaced by a pebble-\nfa\
with no more than $5\cdot f(3m)$ states recognizing the
complement.

In particular, if $f(n)\le O(n^k)$, that is, if there exists
a polynomial transformation for complementing nondeterministic
classical two-way automata, then there must also exist
a polynomial transformation, with the same degree of the
polynomial, for complementing nondeterministic two-way automata
equipped with a pebble.
\etheorem
The argument is very similar to the proof of Theorem~\ref{t:det},
using Theorems~\ref{t:up}(a) and~\ref{t:dw}(a) instead of
Theorems~\ref{t:up}(a) and~\ref{t:dw}(b).

\bcorollary{c:com}
For each pebble-\dfa\ with $m$~states, there exists
a pebble-\dfa\ with at most $60\cdot m$ states recognizing the
complement of the original language.
\ecorollary
This time we use Theorems~\ref{t:up}(b) and~\ref{t:dw}(b),
together with the fact that an $n$-state \dfa\ can be complemented
with no more than $4\cdot n$ states~\cite{GMP07}.

\smallbreak
It was known that a pebble-\dfa\ can be made halting on every
input, and hence a machine for the complement can be obtained by
exchanging accepting with rejecting
states~\cite{CIPR86,CHR91,Si80,Sz94b}. This would give
a pebble-\dfa\ with $O(m\cdots^{2})$ states, where $m$~is the
original number of states and~$s$ the size of the input alphabet.
This way, a linear upper bound is obtained for languages over
a fixed input alphabet, but not in the general case, where the
alphabet size~$s$ can grow exponentially in~$m$ (see,
e.\,g.,~\cite{Si79}). The construction using
Corollary~\ref{c:com} does not depend on the size of the input
alphabet. (However, we are convinced that the upper bound
$60\cdot m$ is quite high and can be improved.)

\bsection{Conclusion}{s:con}
Already in~1978, it was conjectured by Sakoda and
Sipser~\cite{SS78} that there must exist an exponential blow-up,
in the number of states, for the transformation of the classical
\nfa's into \dfa's. Nevertheless, this problem is still open. We
have shown, by Theorem~\ref{t:det} above, that such blow-up could
possibly be derived by proving an exponential gap between
pebble-\nfa's and pebble-\dfa's. Even showing a less impressive
lower bound for the pebble-\nfa\ versus pebble-\dfa\ trade-off,
say, $\Omega(n^k)$ with some $k\ge 3$, would imply the same lower
bound $\Omega(n^k)$ for the classical \nfa\ versus \dfa\
conversion. (To the best of authors' knowledge, the highest lower
bound obtained so far is~$\Omega(n^2)$~\cite{Ch86}.) Since
a pebble automaton is a different computational model, the
argument might use some different witness languages.

Similarly, by Theorem~\ref{t:com}, proving an exponential gap for
the complementation of the pebble-\nfa's would imply the same
exponential gap for the complementation of the classical
\nfa's. This, in turn, would imply the exponential gap for the
trade-off between \nfa's and \dfa's, and also between
pebble-\nfa's and pebble-\dfa's, since the complementation for the
deterministic two-way machines is linear (namely, $4\cdot n$
states for \dfa's, by~\cite{GMP07}, and at most $60\cdot n$
states for pebble-\dfa's, by Corollary~\ref{c:com} above).

\smallbreak
The most natural related open problem is whether the translation
results presented in Theorems~\ref{t:det} and~\ref{t:com} cannot
be generalized to two-way automata equipped with more than one
pebble placed on the input tape. More precisely, we do not know
whether a polynomial trade-off between nondeterministic and
deterministic $k$-pebble two-way automata implies the polynomial
trade-off for automata equipped with $k+1$ pebbles. The same
question can be asked about complementation of multi-pebble
\nfa's. (The argument might be quite difficult, since such
machines can accept nonregular languages: as an example, already
with $2$~pebbles we can easily recognize
$L=\set{a^{n}b^{n}c^{n}:n\ge 0}$.) Nevertheless, the answers to
these questions might bring a deep insight into the world of
$O(\log n)$ space bounded computations, since the multi-pebble
\nfa's and \dfa's correspond to the complexity classes
$\nsp(\log n)$ and $\dsp(\log n)$, respectively (see Section~3.2
in~\cite{Sz94b}). For example, we know that $\nsp(\log n)$ is
closed under complement, by the inductive
counting~\cite{Im88,Szl88}, but the inductive counting technique
increases the number of pebbles.

\smallbreak
Similarly, we need translation(s) among other computational models
that are weak enough to stay within the class of regular languages
but strong enough to provide, in some cases, a more succinct
representation than the classical models. (As an example, we do
not know too much about complexity of two-way automata equipped
with a pushdown of a constant height~\cite{GMP08}.)

\bibliographystyle{eptcs}
\bibliography{geffert}

\end{document}